\documentclass{elsart}
\usepackage{amssymb}
\usepackage{epsfig}
\usepackage{color}
\makeatletter
\renewcommand{\@cite}[1]{#1}
\makeatother
\def\citeb#1{[\cite{#1}]}
\makeatletter
\newdimen\z@ \z@=0pt 
\newskip\z@skip \z@skip=0pt plus0pt minus0pt
\def\m@th{\mathsurround=\z@}
\def\ialign{\everycr{}\tabskip\z@skip\halign} 
\def\eqalign#1{\null\,\vcenter{\openup\jot\m@th
  \ialign{\strut\hfil$\displaystyle{##}$&$\displaystyle{{}##}$\hfil
      \crcr#1\crcr}}\,}
\makeatother
\newcommand{\aff}[2]{Dipartimento di Fisica dell'Universit\`a #1 e 
Sezione INFN, #2, Italy.}
\newcommand{\affd}[1]{Dipartimento di Fisica dell'Universit\`a e 
Sezione INFN, #1, Italy.}
\begin{document}
\begin{frontmatter}

\title{Upper Limit on the $\eta\to\gamma\gamma\gamma$ Branching Ratio with
  the KLOE Detector}

\vskip -1cm 

\collab{The KLOE Collaboration}

\author[Na] {A.~Aloisio},
\author[Na]{F.~Ambrosino},
\author[Frascati]{A.~Antonelli},
\author[Frascati]{M.~Antonelli},
\author[Roma3]{C.~Bacci},
\author[Frascati]{G.~Bencivenni},
\author[Frascati]{S.~Bertolucci},
\author[Roma1]{C.~Bini},
\author[Frascati]{C.~Bloise},
\author[Roma1]{V.~Bocci},
\author[Frascati]{F.~Bossi},
\author[Roma3]{P.~Branchini},
\author[Moscow]{S.~A.~Bulychjov},
\author[Roma1]{R.~Caloi},
\author[Frascati]{P.~Campana},
\author[Frascati]{G.~Capon},
\author[Na]{T.~Capussela},
\author[Roma2]{G.~Carboni},
\author[Roma3]{F.~Ceradini},
\author[Pisa]{F.~Cervelli},
\author[Na]{F.~Cevenini},
\author[Na]{G.~Chiefari},
\author[Frascati]{P.~Ciambrone},
\author[Virginia]{S.~Conetti},
\author[Roma1]{E.~De~Lucia},
\author[Frascati]{P.~De~Simone},
\author[Roma1]{G.~De~Zorzi},
\author[Frascati]{S.~Dell'Agnello},
\author[Karlsruhe]{A.~Denig},
\author[Roma1]{A.~Di~Domenico},
\author[Na]{C.~Di~Donato},
\author[Pisa]{S.~Di~Falco},
\author[Roma3]{B.~Di~Micco}\footnote{Corresponding author: B.~Di Micco, Universit\`a
  ``Roma Tre'', Via della Vasca Navale, 84,
 I-00146, Roma, Italy, e-mail dimicco@fis.uniroma3.it},
\author[Na]{A.~Doria},
\author[Frascati]{M.~Dreucci},
\author[Bari]{O.~Erriquez},
\author[Roma3]{A.~Farilla},
\author[Frascati]{G.~Felici},
\author[Roma3]{A.~Ferrari},
\author[Frascati]{M.~L.~Ferrer},
\author[Frascati]{G.~Finocchiaro},
\author[Frascati]{C.~Forti},
\author[Roma1]{P.~Franzini},
\author[Roma1]{C.~Gatti},
\author[Roma1]{P.~Gauzzi},
\author[Frascati]{S.~Giovannella},
\author[Lecce]{E.~Gorini},
\author[Roma3]{E.~Graziani},
\author[Pisa]{M.~Incagli},
\author[Karlsruhe]{W.~Kluge},
\author[Moscow]{V.~Kulikov},
\author[Roma1]{F.~Lacava},
\author[Frascati]{G.~Lanfranchi}
\author[Frascati,StonyBrook]{J.~Lee-Franzini},
\author[Roma1]{D.~Leone},
\author[Frascati,Beijing]{F.~Lu},
\author[Frascati]{M.~Martemianov},
\author[Frascati]{M.~Matsyuk},
\author[Frascati]{W.~Mei},
\author[Na]{L.~Merola},
\author[Roma2]{R.~Messi},
\author[Frascati]{S.~Miscetti},
\author[Frascati]{M.~Moulson},
\author[Karlsruhe]{S.~M\"uller},
\author[Frascati]{F.~Murtas},
\author[Na]{M.~Napolitano},
\author[Roma3]{F.~Nguyen},
\author[Frascati]{M.~Palutan},
\author[Roma1]{E.~Pasqualucci},
\author[Frascati]{L.~Passalacqua},
\author[Roma3]{A.~Passeri},
\author[Frascati,Energ]{V.~Patera},
\author[Na]{F.~Perfetto},
\author[Roma1]{E.~Petrolo},
\author[Roma1]{L.~Pontecorvo},
\author[Lecce]{M.~Primavera},
\author[Frascati]{P.~Santangelo},
\author[Roma2]{E.~Santovetti},
\author[Na]{G.~Saracino},
\author[StonyBrook]{R.~D.~Schamberger},
\author[Frascati]{B.~Sciascia},
\author[Frascati,Energ]{A.~Sciubba},
\author[Pisa]{F.~Scuri},
\author[Frascati]{I.~Sfiligoi},
\author[Frascati,Budker]{A.~Sibidanov},
\author[Frascati]{T.~Spadaro},
\author[Roma3]{E.~Spiriti},
\author[Roma1]{M.~Testa},
\author[Roma3]{L.~Tortora},
\author[Frascati]{P.~Valente},
\author[Karlsruhe]{B.~Valeriani},
\author[Pisa]{G.~Venanzoni},
\author[Roma1]{S.~Veneziano},
\author[Lecce]{A.~Ventura},
\author[Roma1]{S.Ventura},
\author[Roma3]{R.Versaci},
\author[Na]{I.~Villella},
\author[Frascati,Beijing]{G.~Xu}

\clearpage
\address[Bari]{\affd{Bari}}
\address[Beijing]{Permanent address: Institute of High Energy 
Physics, CAS,  Beijing, China.}
\address[Frascati]{Laboratori Nazionali di Frascati dell'INFN, 
Frascati, Italy.}
\address[Karlsruhe]{Institut f\"ur Experimentelle Kernphysik, 
Universit\"at Karlsruhe, Germany.}
\address[Lecce]{\affd{Lecce}}
\address[Moscow]{Permanent address: Institute for Theoretical 
and Experimental Physics, Moscow, Russia.}
\address[Na]{Dipartimento di Scienze Fisiche dell'Universit\`a 
``Federico II'' e Sezione INFN,
Napoli, Italy}
\address[Budker]{Permanent address: Budker Institute of Nuclear Physics, 
Novosibirsk, Russia}
\address[Pisa]{\affd{Pisa}}
\address[Energ]{Dipartimento di Energetica dell'Universit\`a 
``La Sapienza'', Roma, Italy.}
\address[Roma1]{\aff{``La Sapienza''}{Roma}}
\address[Roma2]{\aff{``Tor Vergata''}{Roma}}
\address[Roma3]{\aff{``Roma Tre''}{Roma}}
\address[StonyBrook]{Physics Department, State University of New 
York at Stony Brook, USA.}
\address[Virginia]{Physics Department, University of Virginia, USA.}

\begin{abstract}
\noindent We have searched for the $C$-violating decay 
$\eta\to\gamma\gamma\gamma$ in a sample of $\sim$18 million $\eta$ 
mesons produced in $\phi \to \eta \gamma$ decays, collected with the 
KLOE detector at the Frascati $\phi$-factory DA$\Phi$NE. No signal is 
observed and we obtain the upper limit 
BR$(\eta\to\gamma\gamma\gamma)\le1.6\times 10^{-5}$ at 90\% CL.
\end{abstract}
\end{frontmatter}

\let\cl=\centerline
\def\figbox#1;#2;{\parbox{#2cm}{\epsfig{file=#1.eps,width=#2cm}}}
\def\figboxc#1;#2;{\cl{\figbox#1;#2;}}
\def\ie{{\it i.\kern-.5pt e.\kern1pt}}  \def\etal{{\it et al.}}
\def\up#1{$^{#1}$}  \def\dn#1{$_{#1}$}
\def\ifm#1{\relax\ifmmode#1\else$#1$\fi}
\def\deg{\ifm{^\circ}}  \def\epm{\ifm{e^+e^-}}
\def\ff{$\phi$--factory}  \def\DAF{DA$\Phi$NE}  \def\f{\ifm{\phi}} 
\def\po{\ifm{\pi^0}}  \def\pio{\ifm{\pi^0\pi^0}}
\def\gam{\ifm{\gamma}} 
\def\ab{\ifm{\sim}}  \def\x{\ifm{\times}}
\def\pt#1,#2,{\ifm{#1\x10^{#2}}}
\def\minus{$-$}  
\def\to{\ifm{\rightarrow}} \def\ett{\ifm{\eta}}
\definecolor{myblue}{rgb}{0,0,.7}
\def\red{\color{red}} \def\blue{\color{myblue}} \def\bk{\color{black}}
\def\mag{\color{magenta}}

The decay \ett\to\gam\gam\gam\ is forbidden by charge-conjugation 
invariance, if the weak interaction is ignored. The present limit for the 
$\eta\to3\gam$ branching ratio, $BR(\eta\to3\gam)\le5 \times10^{-4}$ at 
95\% CL, is based on the result of the GAMS2000 experiment at Serpukhov 
\citeb{GAMS2000}, which studied neutral decays of $\eta$ mesons from the 
reaction $\pi ^{-}p\rightarrow\eta n$ at a beam momentum of 30 GeV/c.

We have searched with KLOE for the decay \ett\to\gam\gam\gam\ among 
four-photon events, corresponding to the two step process 
\f\to\ett\gam, \ett\to\gam\gam\gam. The KLOE detector 
[\cite{K-dc}--\cite{K-trigger}], operates at the Frascati \epm\ collider 
\DAF\ \citeb{Dafne}, which runs at a CM energy $W$ equal to the \f-meson 
mass, $W$\ab1019.5 MeV. Copious $\eta$-meson production is available 
from the decay $\phi\to\eta\gamma$, with a branching ratio of 1.3\%. 
The highest $\phi$-production rate that has been obtained to date 
was \ab240 $\phi$/s, corresponding to \ab3.1 \ett/s, in October 2002.
At \DAF, because of the beam-crossing angle,
$\phi$ mesons are 
produced with a small transverse momentum, 12.5 MeV/$c$, in the horizontal 
plane. The present analysis is based on data collected in the years 
2001 and 2002 for an integrated luminosity of 410 pb$^{-1}$, 
corresponding to \pt1.8,7, \ett\ mesons produced.

The KLOE detector consists of a large cylindrical drift chamber
\citeb{K-dc}, DC, surrounded by a lead/scintillating-fiber sampling
calorimeter \citeb{K-emc}, EMC, both immersed in a solenoidal
magnetic field of 0.52 T with the axis parallel to the beams. Two small
calorimeters \citeb{K-qcal} are wrapped around the quadrupo\-les of the  
low-$\beta$ insertion to complete the detector hermeticity. The
DC tracking volume extends from 28.5 to 190.5 cm in radius and is
330 cm long, centered around the interaction point. The DC momentum 
resolution for charged particles is $\delta p_\perp/p_\perp$=0.4\%. 
Vertices are reconstructed with an accuracy of 3 mm. The calorimeter
is divided into a barrel and two endcaps, and covers 98$\%$ of the 
total
solid angle. Photon energies and arrival times are measured with
resolutions $\sigma_{E}/E = 0.057/{\sqrt{E \ ({\rm GeV})}}$  and
$\sigma_{t} = 54 \ {\rm ps} /{ \sqrt{E \ ({\rm GeV})}} \oplus 50 \
{\rm ps}$, respectively. Photon-shower centroid positions are measured 
with an accuracy of $\sigma=1\ {\rm cm}/\sqrt{E\ ({\rm GeV})}$ along the 
fibers, and 1 cm in the transverse direction.
A photon is defined as a cluster of energy deposits in the calorimeter 
elements that is not associated to a charged particle. We require
the distance between the cluster centroid and the nearest entry point of 
extrapolated tracks be greater than 3\x$\sigma(z,\phi)$.

The trigger \citeb{K-trigger} uses information from both the
calorimeter and the drift chamber. The EMC trigger requires two
local energy deposits above threshold ($E\!>\!50$ MeV in the barrel,
$E\!>\!150$ MeV in the endcaps). Recognition and rejection of
cosmic-ray events is also performed at the trigger level by checking
for the presence of two energy deposits above 30 MeV in the
outermost calorimeter planes. The DC trigger is based on the
multiplicity and topology of the hits in the drift cells. The
trigger has a large time spread  with respect to the time distance
between consecutive beam crossings. It is however synchronized
with the machine radio frequency divided by four, $T_{\rm sync}$=10.85 
ns, with an accuracy of 50 ps. For the  2001-2002 data
taking, the bunch crossing period was $T$=5.43 ns. The time ($T_0$) of
the bunch crossing producing an event is determined offline during
event reconstruction.

The sensitivity of the search for  $\eta$\to\gam\gam\gam\ in KLOE is 
largely dominated by the ability to reject background. The dominant 
process producing four photons is $e^+e^-\to\omega\gam$, due to 
initial-state radiation of a hard photon, followed by 
$\omega\!\to\!\pi^0(\to\!2\gam)\gam$. Other processes with neutral 
secondaries only are also relevant. They can mimic four-photon events because 
of the loss of photons, addition of photons from machine background, or 
photon shower splitting. All the above effects are very difficult to 
reproduce accurately with Monte Carlo (MC) simulation. We therefore base our 
background estimates on data, and use the MC only to 
evaluate the efficiency. An $\eta$\to3\gam\ generator using 
phase space for the internal variable distribution in the three-body decay 
has been used to produce 120,000 $\phi \to \gamma\eta$, 
$\eta\to3\gamma$ events.

For the analysis, only events without charged particle tracks are 
considered. The central value of the position of the beam-interaction 
point (IP), the CM energy, and the transverse momentum of the $\phi$ 
are obtained 
run by run from large samples of Bhabha scattering events.
The following requirements have been used to isolate \f\to4\gam\ 
candidates:
\begin{enumerate}
\item The four photons must have
\begin{itemize}
\item  reconstructed velocity consistent with the speed of light, 
$|t-r/c|<5\sigma_{t}$, where $r$ is the distance traveled, $t$ is the 
time of flight and $\sigma _t$ is the time resolution;
\item photon energy $E_\gamma>$50 MeV;
\item photon polar angle $\theta>$24.5\deg.
\end{itemize}
\item The total energy and momentum of the four prompt photons must satisfy 
$\Sigma_i E_i\!>$800 MeV and $|\Sigma_i\vec p_i|<$200 MeV/$c$;
\item The opening angle between any photon pair must satisfy 
$\theta_{\gamma \gamma}>15^\circ$.
\end{enumerate}

83,906 events pass the cuts above. A kinematic fit is used to improve 
the energy-momentum resolution. The input variables $x_i$ of the fit are
\begin{itemize}
\item[-] the coordinates of the photon clusters in the 
calorimeter;
\item[-] the energies of the clusters;
\item[-] the times of flight of the photons;
\item[-] the coordinates of the $e^+e^-$ interaction 
point; 
\item[-] the energy and momentum of the $\phi$ meson.
\end{itemize}
We minimize the $\chi ^2$ function
$$\chi^2=\sum_{i}{\frac{(x_i-\mu_i)^2}{\sigma_i^2}}+\sum_{j}{\lambda_j 
F_j(\mu_k)},$$
where $F_j(\mu_k)$ are the energy, momentum, and time constraints and 
$\lambda_i$ are Lagrangian multipliers.
The $\chi ^2$ value of the fit is used to reject background. Events 
with $\chi^2<25$ are retained, the number of degrees of freedom 
being 8.

After this cut we are left with 52577 events. The residual background after the cut is due 
to events with  neutral pions (fig.~\ref{fig:mp0g},
left), coming mainly from $e^+e^-
\to\omega\gamma$ with $\omega\to\po
\gamma$. This can be seen in fig.~\ref{fig:mp0g}, right, where the invariant mass of the $\pi^{0}$ and the highest energy 
photon, in the $\pi^0 \gamma \gamma$ hypothesis, \gam\dn{\rm hi}, shows a clear peak at the $\omega$ mass.
\begin{figure}[hbtp]
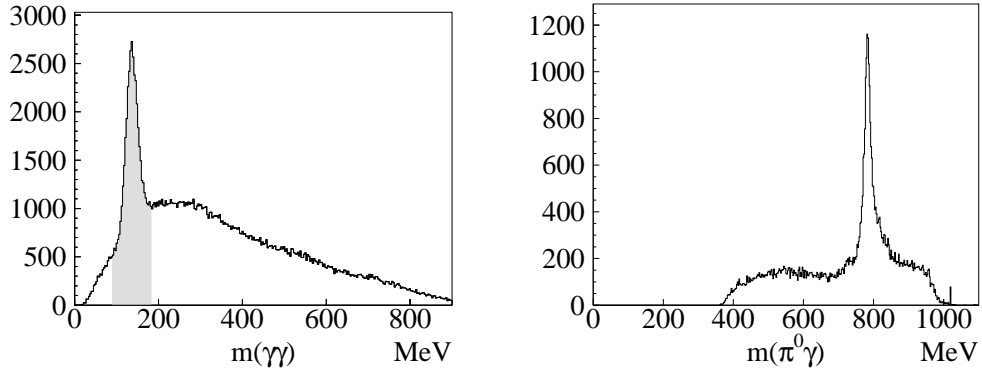

\kern1cm\figbox mp0;6;\kern1cm\figbox mp0g1;6;
\caption{Distribution of the invariant masses  $m(\gam\gam)$ computed for all photon pairs, left and $m(\pi^0 \gam_{\rm hi})$, right. The shaded 
interval is removed before further analysis.}
\label{fig:mp0g}
\end{figure}
Other background sources with a $\pi^0$ in the final state are the 
decays \f\to\po\gam, $\phi\to f_0\gam\to\pio\gam$, and $\phi\to a_0\gam\to\eta
\pi^0 \gamma$. We reject the main part of these events by a cut on the invariant mass of 
any photon pairs: $90<m(\gamma \gamma)< 180$ MeV.

8,268 events survive the cuts. In the decay \f\to\ett\gam, the energy of
the recoil photon in the  CM of the \f\ is 363 MeV. In the complete chain 
\f\to\ett\gam, \ett\to3\gam, 363 MeV is also the most probable energy 
of the most energetic photon, \gam\dn{\rm hi}. Fig. 
\ref{fig:maxenergy}, left, shows an MC simulation of the \gam\dn{\rm hi} 
energy spectrum for the signal. Fig. \ref{fig:maxenergy}, right, shows the 
$E(\gam_{\rm hi})$ distribution for the data sample. No peak is 
observed around 363 MeV.
\begin{figure}[hbtp]
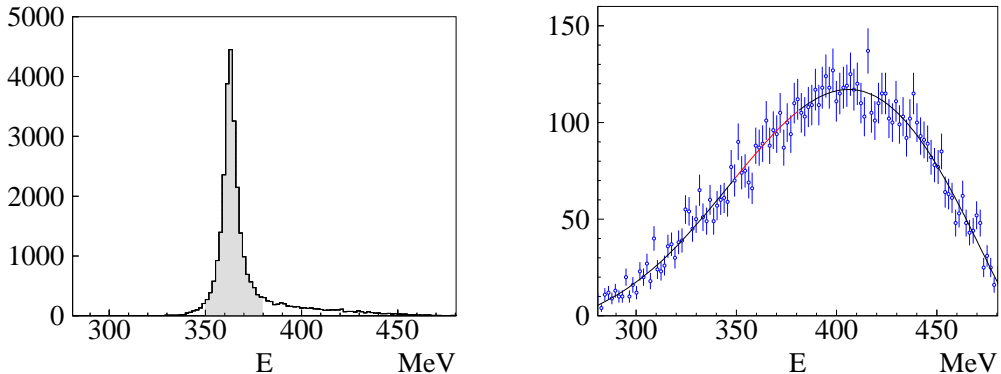

\cl{\figbox emaxfitdraftmc;6;\kern12mm \figbox emaxfitdraftdata;6;}
\caption{Distribution of the energy $E(\gam_{\rm hi})$, in the $\phi$ 
CM, for the MC simulated signal (left) and for the data (right). See text 
for discussion of the background fit.}\label{fig:maxenergy}
\end{figure}
To evaluate an upper limit on the number of $\eta \to 3 \gamma$
events, we choose as the signal region the interval 350$<\!E(\gam_{\rm 
hi})\!<$379.75 MeV (17 bins, 1.75-MeV wide). We estimate the background 
by fitting the $E(\gam_{\rm hi})$ distribution on both sides of the 
expected signal region, in the intervals 280$<E<$350 and 
379.75$<E<$481.25 MeV.
We fit the background using 3\up{\rm rd} to 6\up{\rm th} order 
polynomials. The $5^{\rm th}$ order polynomial shown in fig. 
\ref{fig:maxenergy} gives the best fit, 
with $\chi^2/\textrm{dof}=78/92=0.85$. We use the result to obtain the 
expected number of background events in each bin, $N^b_i$. The total
number of observed events in the signal window is 1513 while 
from integration of the polynomial we obtain 1518 events in the same region.

The upper limits have been evaluated using Neyman's construction 
procedure [\cite{FeldmannCousins}]. 
To evaluate the agreement with the background distribution in the 
signal region, we use
$$F = \sum_i \frac{(N_i - N^b_i)^2}{N^b_i},$$
where $N_i$ is the number of observed counts in the $i^\mathrm{th}$ bin, and
the sum is over bins in the signal region.
We obtain the distribution function for $F$ for various values of the
number of signal counts $s$ as follows.
First, we \emph{construct} the values $N_i$ 
by sampling a Poisson distribution with mean $\left< N_i(s) \right> = N_i^b + s\times f_i$, 
where $f_i$ is the fraction of signal events
(fig. \ref{fig:maxenergy}, left) 
in the $i^{th}$ bin, and evaluate $F$.
Repeating this procedure $10^6$ times for each value of $s$ then gives
the complete p.d.f., which is numerically integrated to obtain
the 90\% and 95\% contours in Neyman's construction.
We then evaluate $F$ using the \emph{observed} $N_i$. We find
$F=13.45$, from which we obtain 
$$N_{\eta \to 3 \gamma } \le 63.1 \textrm{ at 90\% CL};\ \ \le 80.8 
\textrm{ at 95\% CL}.$$
To convert this result into an upper limit for the branching
ratio, we normalize to the number of $\eta \to 3 \pi^0$ events 
\citeb{MiscettiGiovannella} found in the same data sample, $N(\eta \to 
3\pi^0)$=2,431,917. 
The efficiencies are $\epsilon(\eta
\to 3\pi^0)\kern-3pt=\kern-3pt0.378 \pm 0.008 (syst.) \pm 0.001 (stat.)$ and 
$\epsilon(\eta\to
3\gamma)$=$\kern1pt0.200 \pm 0.001(stat.) \pm 0.002 (syst.) \pm 0.006 (\chi^2_{cut})$. 
The systematic error includes residual uncertainities on the photon detection efficiency \citeb{Spadaro}. For the ratio of the two branching ratios we
obtain
$$\eqalign{\frac{{\rm BR}(\eta \to 3\gamma)}{{\rm BR}(\eta \to 3 \pi 
^0)}=
 \frac{N_{\eta \to 3 \gamma}~\epsilon_{\eta \to 3 \pi^0}}
 {N_{\eta \to 3 \pi^0}~\epsilon_{\eta \to 3 \gamma}}
 &\le 4.9 \times 10^{-5}\textrm{ 90\% CL},\cr
 \noalign{\vglue-.8cm}\cr
 &\le 6.3 \times 10^{-5}\textrm{ 95\% CL}.\cr}$$
Using the value BR$(\eta \to 3 \pi^0)=(32.51 \pm 0.29)\%$
\citeb{PDG2002}, we derive the upper limit 
\[
BR(\eta\to3\gamma)
\le 1.6 \times 10^{-5} \quad \textrm{at 90\% CL} \quad \textrm{and} \quad \le 2.0 \times 10^{-5} \quad \textrm{at 95\% CL}. 
\]
The efficiency quoted above for \ett\to3\gam, which depends on the cut 
$\chi^2<25$ applied after the kinematic fit of all four-photon events, 
is evaluated by MC simulation. We check the validity of the MC 
result by comparison with the $\chi^2$ distribution for radiative events 
\epm\to\gam$\omega$\to\gam\gam\po\to4\gam. A sample of these events is 
selected from among all four-photon candidates by requiring $128<m(\gam\gam)<145$ 
MeV for the neutral pion and $760<m(\po\gam)<815$ MeV for the $\omega$. 
The fraction of these events with $\chi^2<25$ after the kinematic
 fit differs 
from the MC estimate by \ab3\%. This value is included in the quoted error for $\epsilon(\eta\to3\gamma)$.

To check whether the kinematic fit introduces a bias in the 
energy distribution of the signal photons,
we have analyzed a sample of 
\f\to\ett\gam\dn{\rm rec} \to\gam\gam \gam\dn{\rm rec} events, in 
which the energy of the recoil photon is the same as in the case of interest. 
Fig. \ref{fig:erad2g} shows the energy distribution of the 
photons as obtained after the kinematic fit for data and MC events.
 The two distributions are in good agreement 
within errors.

\begin{figure}[!htbp]
\begin{center}
\includegraphics[scale=0.3]{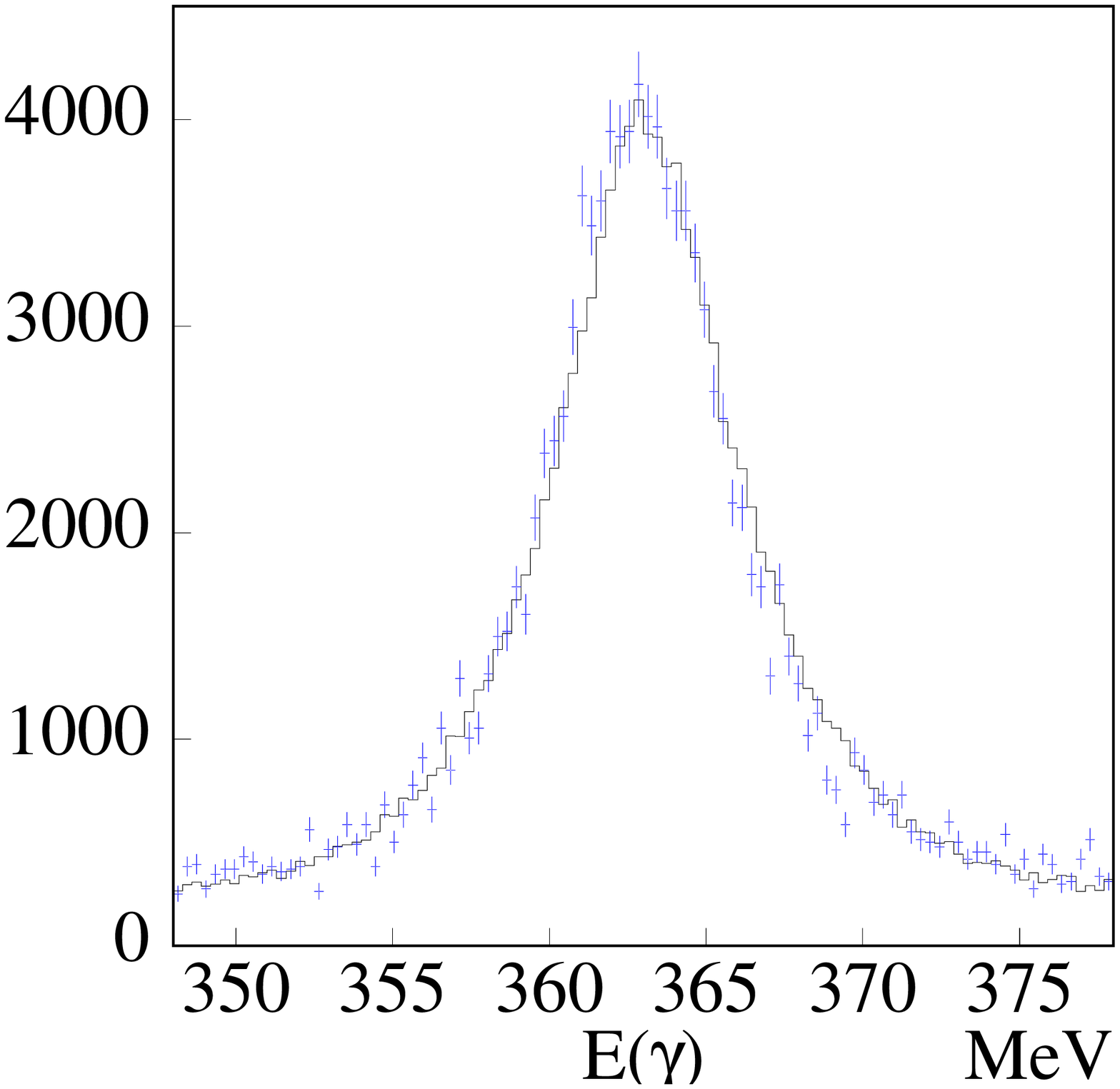}
\end{center}
\caption{Distribution of $E(\gam)$ for the 
$\phi \to \eta \gamma \to (\gamma \gamma) \gamma$ event sample for the 
data (points) and MC simulated events (continuous histogram).}
\label{fig:erad2g}
\end{figure}

The stability of the upper limit versus the background 
estimate has been checked by comparing the results of polynomials of 
different degree 
 for fitting the $E(\gam_{\rm hi})$ distribution outside the 
signal region. A $3^{\rm rd}$ order polynomial doesn't describe 
the background shape well. A $4^{\rm th}$ order polynomial gives a lower
value for the signal yield, while a $6^{\rm th}$ order polynomial gives 
the same result. We have also checked the stability of the result 
by changing the window chosen for evaluation of the upper limit
obtaining a maximum variation of 11\%.
We have also evaluated the $\eta \to 3 \gamma$ acceptance using the matrix element of ref.[\cite{dicus}] and we find
a value 5\% lower. Therefore systematic effects can be summarized: background estimation and window variation 11\%,
$\epsilon(\eta \to 3\pi^0)/\epsilon(\eta \to 3 \gamma)$ 1\%, $\chi^2$ cut 3\%, decay model 5\%.
We thus feel confident about the procedure adopted. 
Our limit
$${\rm BR}(\eta \to \gamma \gamma \gamma) \le 1.6 \times 10^{-5}\textrm{ at 
90\% CL}  \ {\rm or}\ \leq 2.0 \times 10 ^{-5}\textrm{ at 95\% CL}$$
is the strongest limit at present against possible violation of 
charge-conjugation invariance in the decay \ett\to3\gam.\footnote{A 
preliminary unpublished 2002 result by the Crystal Ball, 
$BR(\eta\to3\gamma)\le1.8 \times10^{-5}$ at 90\% CL, is mentioned in 
\citeb{Nefkens&Price}.}
An estimate for $\Gamma(\eta\to3\gam)$, including contributions from weak 
interactions, is given in Ref. \cite{herc}. Using the estimate for 
$\pi^0\to3\gam$ [\cite{dicus}], one finds BR$(\eta\to3\gam)<10^{-12}$, 
which is quite a long way from the result above. The absence of the decay 
\ett\to3\gam\ therefore confirms the validity of charge-conjugation 
invariance in strong and electromagnetic interactions.

We thank the DA$\Phi$NE team for their efforts in maintaining low
background running conditions and their collaboration during all
data taking. We want to thank our technical staff: G.F. Fortugno
for his dedicated work to ensure efficient operations of the
KLOE Computing Center; M. Anelli for his continuous support of the
gas system and detector safety; A. Balla, M. Gatta, 
G. Corradi and G. Papalino for electronics maintenance;
M. Santoni, G. Paoluzzi and R. Rosellini for general support of
the detector; C. Pinto (Bari), C. Pinto (Lecce), C. Piscitelli and
A. Rossi for their help during major maintenance periods. This work was
supported in part by DOE grant DE-FG-02-97ER41027; by EURODAPHNE,
contract FMRX-CT98-0169; by the German Federal Ministry of
Education and Research (BMBF) contract 06-KA-957; by
Graduiertenkolleg `H.E. Phys. and Part. Astrophys.' of Deutsche
Forschungsgemeinschaft, Contract No. GK 742; by INTAS, contracts
96-624, 99-37; and by TARI, contract HPRI-CT-1999-00088.

\end{document}